\documentclass{aa} 

\usepackage{graphicx}
\usepackage{txfonts}
\usepackage{amsmath,amssymb}
\usepackage{ccaption}

\usepackage[colorlinks=true, allcolors=blue]{hyperref}
\usepackage{color}

\begin{document}

   \title{HCN (1-0) opacity of outflowing gas in Arp 220W}

%   \subtitle{I. Overviewing the $\kappa$-mechanism}

   \author{J. Z. Wang\inst{1,2}, 
          S. Liu\inst{3,4},
          Z-Y. Zhang\inst{5,6},
          \and
          Y. Shi\inst{5,6}
          }

   \institute{Shanghai Astronomical Observatory, Chinese Academy of Sciences,80 Nandan Road, Shanghai, 200030, PR China\\
              \email{jzwang@shao.ac.cn}
         \and
             Key Laboratory of Radio Astronomy, Chinese Academy of Sciences,  10 Yuanhua Road, Nanjing, JiangSu 210033, PR China
         \and
          National Astronomical Observatories, Chinese Academy of Sciences, Beijing 100012, PR China\\
             \email{liushu@nao.cas.cn}
          \and    
             CAS Key Laboratory of FAST, National Astronomical Observatories, Chinese Academy of Sciences, Beijing 100012, PR China
                \and
             School  of Astronomy and Space Science, Nanjing University, Nanjing,  210093, PR China
             \and Key Laboratory of Modern Astronomy and Astrophysics (Nanjing University), Ministry of Education, Nanjing 210093, PR China\\
             }

 \date{Received xx / accepted xx}

% \abstract{}{}{}{}{} 
% 5 {} token are mandatory
 
  \abstract
  % context heading (optional)
  % {} leave it empty if necessary  
   {We present  our findings for the HCN/H$^{13}$CN 1-0 line ratio in the molecular  outflow of Arp 220 west based on high-resolution ALMA data. }
  % aims heading (mandatory)
   {Molecular gas masses in the outflowing gas of galaxies driven by active galactic nuclei (AGNs) or starbursts are important parameters for understanding the feedback of these latter two phenomena and
star-formation quenching. The conversion factor of line luminosities to masses is related to the optical depth of the molecular lines. }
  % methods heading (mandatory)
   { Using H$^{13}$CN 1-0, the isotopic line  of HCN 1-0, to obtain the line ratio of HCN/H$^{13}$CN 1-0 is an ideal way to derive the optical depth of HCN 1-0 in outflowing gas.}
  % results heading (mandatory)
   { With the nondetection of H$^{13}$CN 1-0 in the outflowing gas, a 3$\sigma$ lower limit of HCN/H$^{13}$CN 1-0 line ratio is obtained, which is  at least three times higher than that found in the whole of the whole system of Arp 220.   The high HCN/H$^{13}$CN 1-0 line ratio indicates low opacity of HCN 1-0 in the outflowing gas, even though the upper limit of HCN 1-0 opacity obtained here   is still not good enough to draw any robust conclusions if  HCN 1-0 is  optically thin.  A lower conversion  factor of HCN 1-0 luminosity to dense gas mass in the outflowing gas should be used than that used for the host galaxy of Arp 220.}
  % conclusions heading (optional), leave it empty if necessary 
 {}
 

   \maketitle
%
%-------------------------------------------------------------------

\section{Introduction}

Mass outflows,  a major  manifestation of the   feedback from active galactic nuclei (AGNs) and circumnuclear extreme starbursts (SBs), is one key element for understanding galaxy evolution \citep{2010A&A...518L.155F, 2011ApJ...733L..16S}. Massive molecular outflows driven by AGNs or SBs have been found in  galaxies with CO emissions \citep{2010A&A...518L.155F,2012A&A...543A..99C, 2014A&A...562A..21C,2019MNRAS.483.4586F} 
 and far infrared OH absorption  lines  \citep{2010A&A...518L..41F}. Outflow rates of molecular gas in  starburst-dominated galaxies have been found to be  comparable to or even higher than their star formation rates \citep{2014A&A...562A..21C,2019MNRAS.483.4586F}. Furthermore, it has been suggested that a  high-mass outflow rate can effectively quench star formation, which may cause the host galaxy to quickly leave the star-forming phase to become an early-type gas-poor red galaxy. 

Such outflows can also be detected with dense gas tracers. Examples are HCN,   HCO$^+$, and HNC lines in Mrk 231 \citep{2012A&A...537A..44A,2015A&A...574A..85A},   HCN 1-0 in Arp 220 \citep{2018ApJ...853L..28B},  and HCO$^+$ and HCN 1-0 in NGC 3256 \citep{2018ApJ...868...95M}. The mass-loss rate and  outflow mass can be estimated with line luminosities of HCN and HCO$^+$ wings. However, such estimations strongly depend on the conversion factor from line luminosity to dense gas mass, which is related to the   relative abundance of HCN and HCO$^+$ to hydrogen and  the line opacity. The conversion factor of ${\alpha}_{HCN (1-0)}$=3.3 $M_{\odot}$ (K km s$^{-1}$pc$^{2})^{-1}$  whilst the  optically thick assumption  has been used  to estimate the upper limit of  the mass of the dense molecular outflow in Mrk 231  \citep{2015A&A...574A..85A}, while  ${\alpha}_{HCN (1-0)}$ is 0.24$M_{\odot}$ (K km s$^{-1}$pc$^{2})^{-1}$, if the optically thin assumption is made \citep{2018ApJ...853L..28B}. On the other hand, \cite{2018ApJ...868...95M}  used 0.24 and 10   $M_{\odot}$ (K km s$^{-1}$pc$^{2})^{-1}$ to calculate the lower and upper limit of dense gas mass in NGC 3256.
  Isotopologue lines of HCN and  HCO$^+$ molecules, such as  H$^{13}$CN and  H$^{13}$CO$^+$ lines, can be used to determine the opacities of HCN and  HCO$^+$ lines with line ratios of HCN/H$^{13}$CN and HCO$^+$/H$^{13}$CO$^+$, and were  adopted  to obtain averaged HCN opacities   in nearby galaxies  \citep{2016MNRAS.455.3986W, 2020MNRAS.494.1095L}. 
  
  %However, due to the  weak emission of these lines from outflowing gas, it is really hard to obtain a reliable line ratio for most of galaxies.  

Arp 220, as the  nearest ultraluminous infrared galaxy (ULIRG) in the late stages of a galaxy merger, with strong outflowing gas  at about $\pm$400 km s$^{-1}$ and more than 10$^7M_\odot$ mass  around the western nucleus  (Arp 220W) detected with HCN 1-0  \citep{2018ApJ...853L..28B}, is a good candidate to determine the line ratio of HCN/H$^{13}$CN  1-0 for the outflowing gas, which can be used to determine the  opacity of  HCN 1-0 in outflowing gas. 

In this letter, we describe the  data from  the ALMA archive. First we present our data reduction  in \S2, before presenting our main results   in \S3, and discussions in \S4. We provide a brief summary in \S5.

%--------------------------------------------------------------------
\section{The data}

The data  were taken from the ALMA  archival system (project number: 2015.1.00702.S, PI: L. Barcos-Mu\~{n}oz), the same data set used in \cite{2018ApJ...853L..28B}. Standard  bandpass, phase, and flux calibrations  using scripts from the archive system were done in CASA, as well as  imaging and deconvolution   using natural weighting with  a pixel size of 0.02$''$ and frequency resolution of 3.906 MHz  in the task `tclean'.   The continuum image  obtained from  line-free channels,  the velocity-integrated maps of HCN 1-0  at the rest frequency of   88.6318473 GHz for both red and blue wings, as well as the  velocity integrated maps of the optically thin isotopologue H$^{13}$CN 1-0,  were obtained from the cleaned  datacube (see Figure 1).  Spectra of HCN and H$^{13}$CN 1-0   at the rest frequency of   86.3401764 GHz toward  red and blue wings from the regions marked as dashed cyan  boxes in Figure 1 were also obtained and are presented in Figures 2 and 3.

 \begin{figure}
\includegraphics[width=0.4\textwidth]{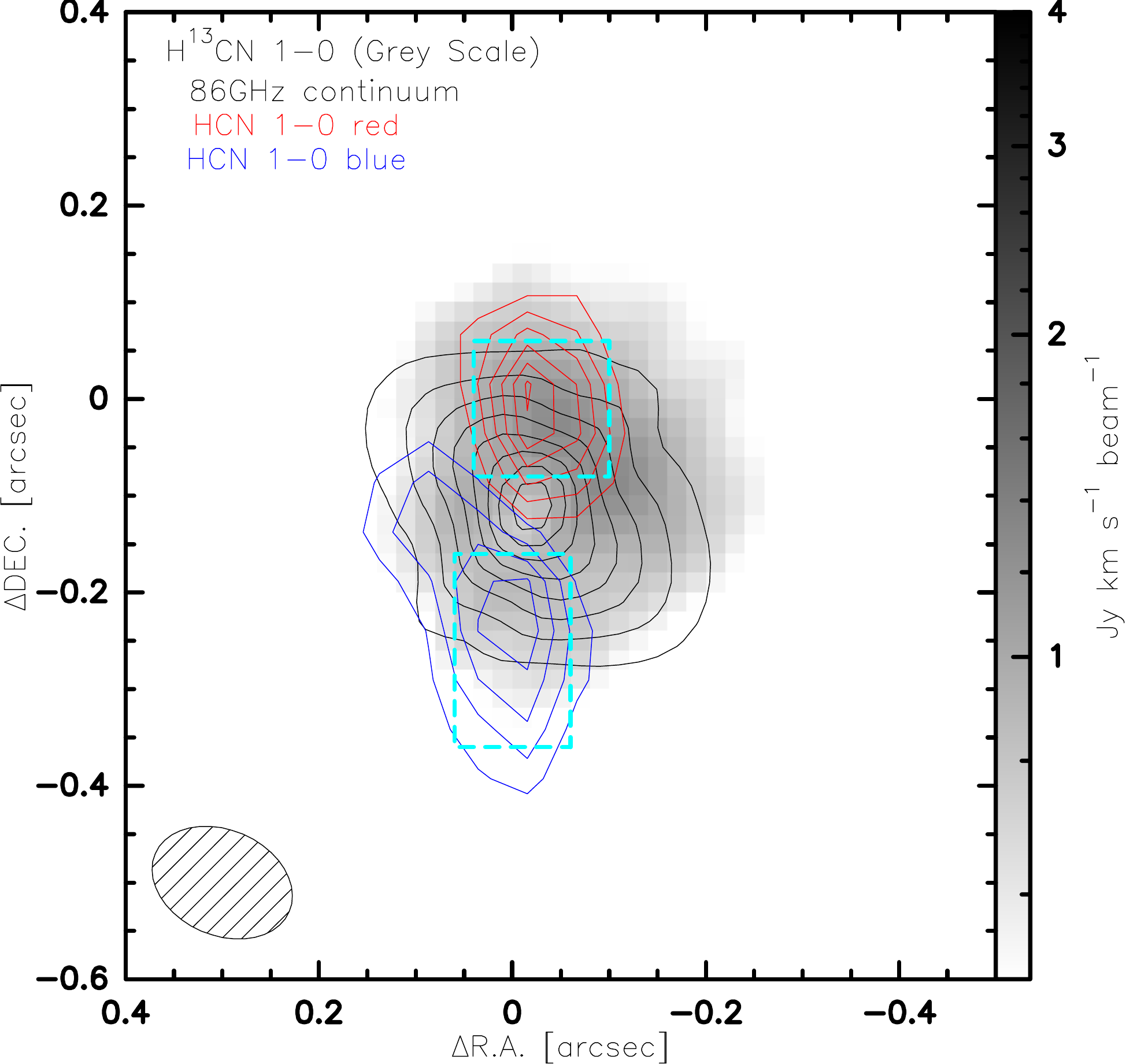}
    \caption{The 86 GHz continuum (black contour  with levels starting  from 3mJy beam$^{-1}$ in steps of 1 mJy beam$^{-1}$), a velocity integrated map of  H$^{13}$CN 1-0 (grey scale, in units of Jy km s$^{-1}$ beam$^{-1}$), a velocity integrated map of  HCN 1-0   red wing (red contour with levels starting  from 0.2 Jy km s$^{-1}$ beam$^{-1}$ in steps of 0.05 Jy km s$^{-1}$ beam$^{-1}$) and a velocity integrated map of  HCN 1-0 blue wing (blue contour  with levels starting  from 0.2 Jy km s$^{-1}$ beam$^{-1}$ in steps of 0.05 Jy km s$^{-1}$ beam$^{-1}$), around the centre of Arp 220 west. The dashed cyan boxes are the regions for the spectra in Figures 2 and 3. The  restoring beam ($0.151''\times0.109"$, $PA=23.4^{\circ}$) of the data cube is shown in the bottom left.   The central coordinates of this map are R.A.: 15:34:57.22 and Dec: 23:30:11.6 (J2000).}
    \label{fig:figure1}
\end{figure}

% Example figure
\begin{figure}
        % To include a figure from a file named example.*
        % Allowable file formats are eps or ps if compiling using latex
        % or pdf, png, jpg if compiling using pdflatex
%       \includegraphics[width=\columnwidth]{example}
\includegraphics[angle=0,scale=.25]{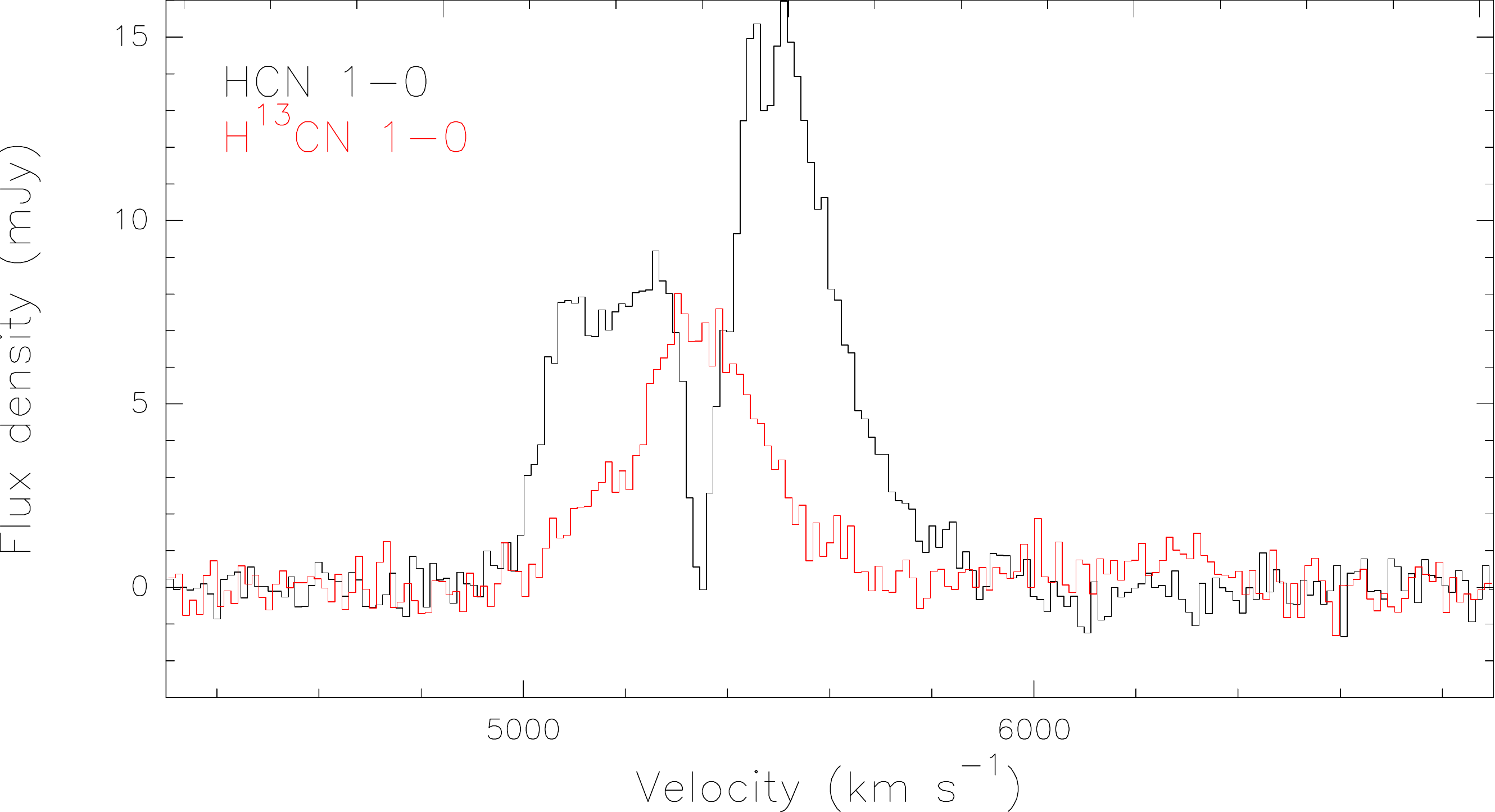}
\includegraphics[angle=0,scale=.25]{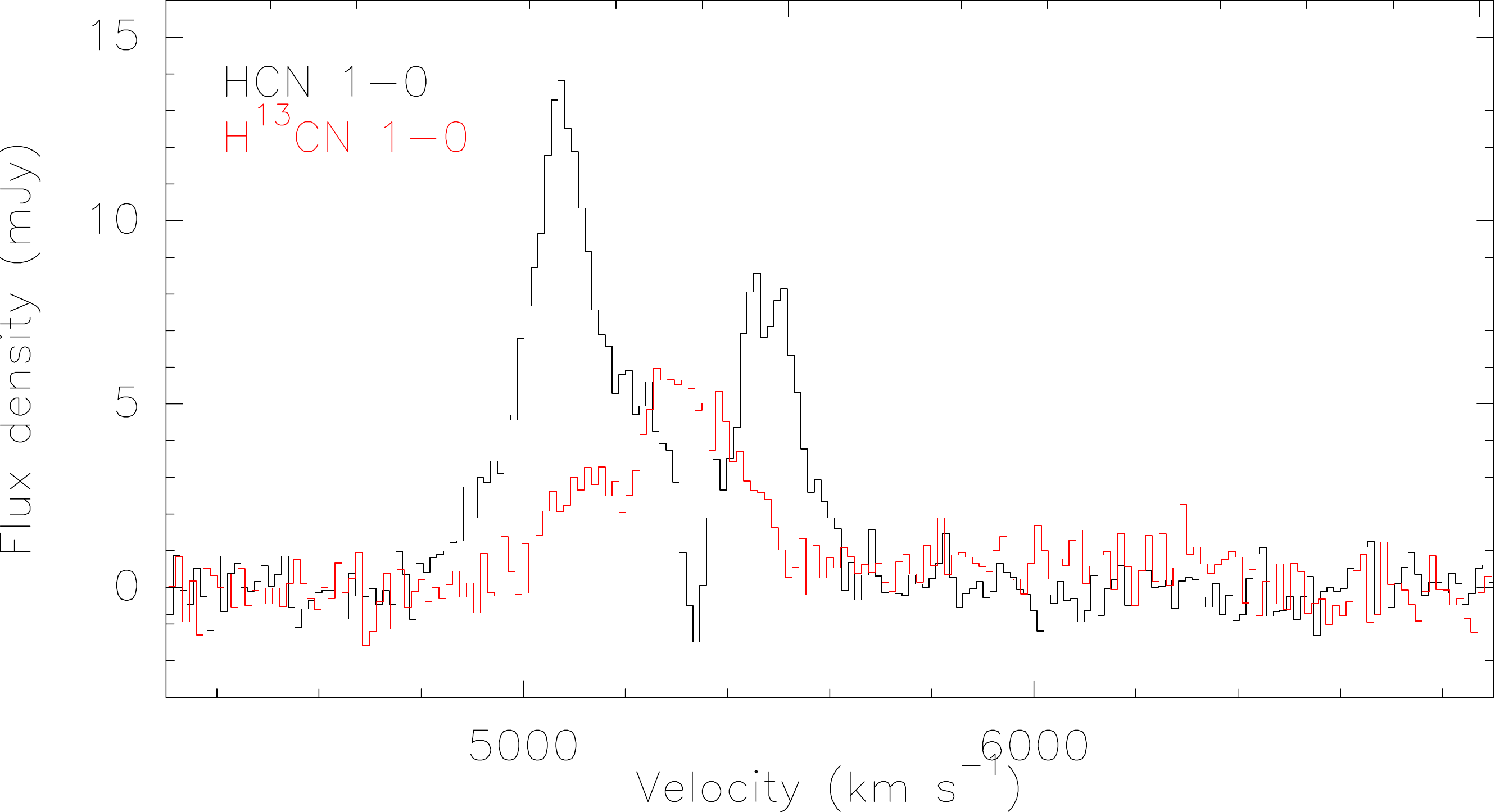}

    \caption{Top: HCN (black) and  H$^{13}$CN 1-0 (red) spectra collected from the red wing region of  the cyan box in Figure 1. Bottom: Same as the top spectra but from the blue wing region.  Radio-defined velocities are used, while the system velocity is about 5350 km s$^{-1}$.}
    \label{fig:figure2}
\end{figure}

\section{Results}

\subsection{Spatial distribution of  outflow traced by HCN}

The velocity-integrated flux distribution of the red and blue wings of HCN 1-0, as well as  the continuum emission around 86 GHz  and the velocity-integrated flux of  H$^{13}$CN 1-0 near the centre of Arp 220W are presented in Figure 1.  Outflows near the centre of Arp 220W   can be seen  with  HCN 1-0 emission, which is consistent with the findings reported  by \cite{2018ApJ...853L..28B}, with a blue wing at the northern side and a red wing at the southern side of the centre of Arp 220W traced by continuum emission.    As the abundance of  H$^{13}$CN is less than one-tenth of  HCN,  H$^{13}$CN 1-0 is normally optically thin in Galactic star forming regions   \citep{2011A&A...534A..77P, 2018A&A...609A.129C} and galaxies  \citep{1998A&A...329..443H,2020MNRAS.494.1095L}. However, H$^{13}$CN 1-0  in Arp 220W shows an offset of the peak emission from that of the continuum, obtained from (see Figure 1). Such an offset may be caused by absorption of   H$^{13}$CN 1-0 towards  continuum emission around the centre region, where the absorption of HCN 1-0 is significantly below the continuum level at some velocity ranges.    HCN 1-0 absorption can be important even at the positions with less continuum emission than the centre (see Figure 2 and 3). 
 
Collecting  the most significant emitting regions of red and blue wings marked as dashed cyan  boxes in Figure 1, HCN and H$^{13}$CN 1-0 spectra are obtained to determine the line ratios of   HCN/H$^{13}$CN 1-0 in the line wings. The red wing displayed in Figure 1 shown with red contours  is integrated from 5620 to 5800 km s$^{-1}$ as radio-defined velocity, while the blue wing shown with blue contours is integrated from 4860 to 5040 km s$^{-1}$.  The SO line at the rest frequency of 86.093983 GHz and HC$^{15}$N 1-0 at the rest frequency of 86.054967 GHz can contaminate the red wing of  H$^{13}$CN 1-0 (see Figure 3), especially with velocity higher than 6000 km s$^{-1}$.  Therefore, in order  to avoid such contaminations when estimating the HCN/H$^{13}$CN1-0 ratio,  even though there are no such contaminations for the HCN 1-0 line,  only the velocity range  between 5620 and 5800 km s$^{-1}$  is integrated for the red wing.

\subsection{HCN/H$^{13}$CN 1-0 line ratio}

 The HCN 1-0 emission is no more than five times  the corresponding H$^{13}$CN 1-0  emission in most of the velocity components towards the red and blue wings of  the outflowing gas  (see Figure 3).  The HCN/H$^{13}$CN  1-0 ratio in the entire system of Arp 220 is about five according to single-dish observations with IRAM 30-meter telescope \citep{2016MNRAS.455.3986W}.   HCN and H$^{13}$CN  1-0 fluxes measured from ALMA data are consistent with those  obtained with the IRAM 30-meter telescope \citep{2016MNRAS.455.3986W}.    HCN/H$^{13}$CN  1-0 ratios towards the outflowing regions  vary along with different velocities (see Figure 2 and 3), from more than five to even less than one at the line centre, which  seems likely to be  caused by  the absorption of HCN 1-0  towards continuum emission or the  self absorption of  HCN 1-0 with a  temperature gradient. Such an absorption, where the velocity of the HCN 1-0 absorption peak is similar to that of the H$^{13}$CN 1-0  emission peak,   will cause an overestimation of opacity if a unique temperature is assumed for the molecular gas.    However, such absorption  is  mainly caused by gas from the disc instead of the outflowing gas, based on the velocity information.  
 
  The line profile of  HCN 1-0   from the red wing region (see Figure 3)  in the blue part from about 4950 to 5100 km s$^{-1}$ agrees well with the spectra of  H$^{13}$CN  1-0  from the same region, with a line ratio of about five. On the other hand,  the intensity of HCN 1-0 is well above five times that of  H$^{13}$CN  1-0 at the red part, especially from 5620 to 5800 km s$^{-1}$.  The SO line at the rest frequency of 86.093983 GHz  contaminates  H$^{13}$CN  1-0 from $\sim$5850 km s$^{-1}$  to 6100 km s$^{-1}$. Thus, it is hard to obtain  a reliable  ratio of HCN and H$^{13}$CN  1-0  at velocity ranges with SO line contamination. SO and HC$^{15}$N 1-0 emission can also be seen in the blue wing regions (See Figure 3). However, the line profiles of  HCN and H$^{13}$CN  1-0 from  the blue wing region agree well with each other in the red part  without   SO  contamination, with a line ratio also similar to five, while HCN 1-0 intensity is well above five times that of  H$^{13}$CN  1-0 in the blue  region (see Figure 3).  In summary,   HCN 1-0 emission is significantly above five times that of H$^{13}$CN  1-0  from the outflowing gas, while it is equal to or  below five times that of the H$^{13}$CN  1-0 emission from the gas that is not outflowing.

The  HCN 1-0 velocity-integrated flux of the red wing in Figures 2 and 3  between 5620 and 5800 km s$^{-1}$ is 631.1$\pm$22.5 mJy km s$^{-1}$, while this value is 794.1$\pm$27.7 mJy km s$^{-1}$ for the blue wing between 4860 and 5040  km s$^{-1}$ (see Table 1). The uncertainties  are estimated with ${\sigma}_{rms}\times\sqrt{{\delta}v{\Delta}V}$, where ${\sigma}_{rms}$ is the rms with baseline fitting for the spectra shown in Figure 2 at the velocity resolution of ${\delta}v$ as $\sim$13.2  km s$^{-1}$, while ${\Delta}V$ is  180 km s$^{-1}$  which is the integrated line width for both red and blue  wings.   The velocity-integrated fluxes of H$^{13}$CN 1-0 at the corresponding velocities are 59.7 and 61.5  mJy km s$^{-1}$  for the red and blue wings, respectively, which is  below the 3$\sigma$ level for both components. The noise level near H$^{13}$CN 1-0 is about 0.94 of that of the  HCN 1-0 line.  Thus, the 3$\sigma$ values 63.2 and 77.9 mJy km s$^{-1}$  are used for the red and blue wings, respectively, which give line ratios of HCN/H$^{13}$CN 1-0  greater than 9.5   and  9.7   in the  red and blue wings, after  converting the units from  mJy for  flux density to mK for brightness temperature. 

 The line ratio   HCN/H$^{13}$CN 1-0 varies  from 7.6$\pm$2.8 in NGC 4418 to 40$\pm$3.6 in M82 within a sample of local galaxies  \citep{2020MNRAS.494.1095L}. Thus, the high HCN/H$^{13}$CN 1-0 ratio  of greater than 9.6 in the line wings of Arp 220W is not extremely high  in galaxies.  On the other hand,  the HCN/H$^{13}$CN 1-0 line ratio of about five for the whole galaxy of Arp 220 \citep{2016MNRAS.455.3986W} indicates that HCN column density in  Arp 220 is extremely high, which is consistent with our knowledge of Arp 220 as a gas-rich late-stage merger with extreme starburst activity.

\begin{figure}
        % To include a figure from a file named example.*
        % Allowable file formats are eps or ps if compiling using latex
        % or pdf, png, jpg if compiling using pdflatex
%       \includegraphics[width=\columnwidth]{example}
\includegraphics[angle=0,scale=.25]{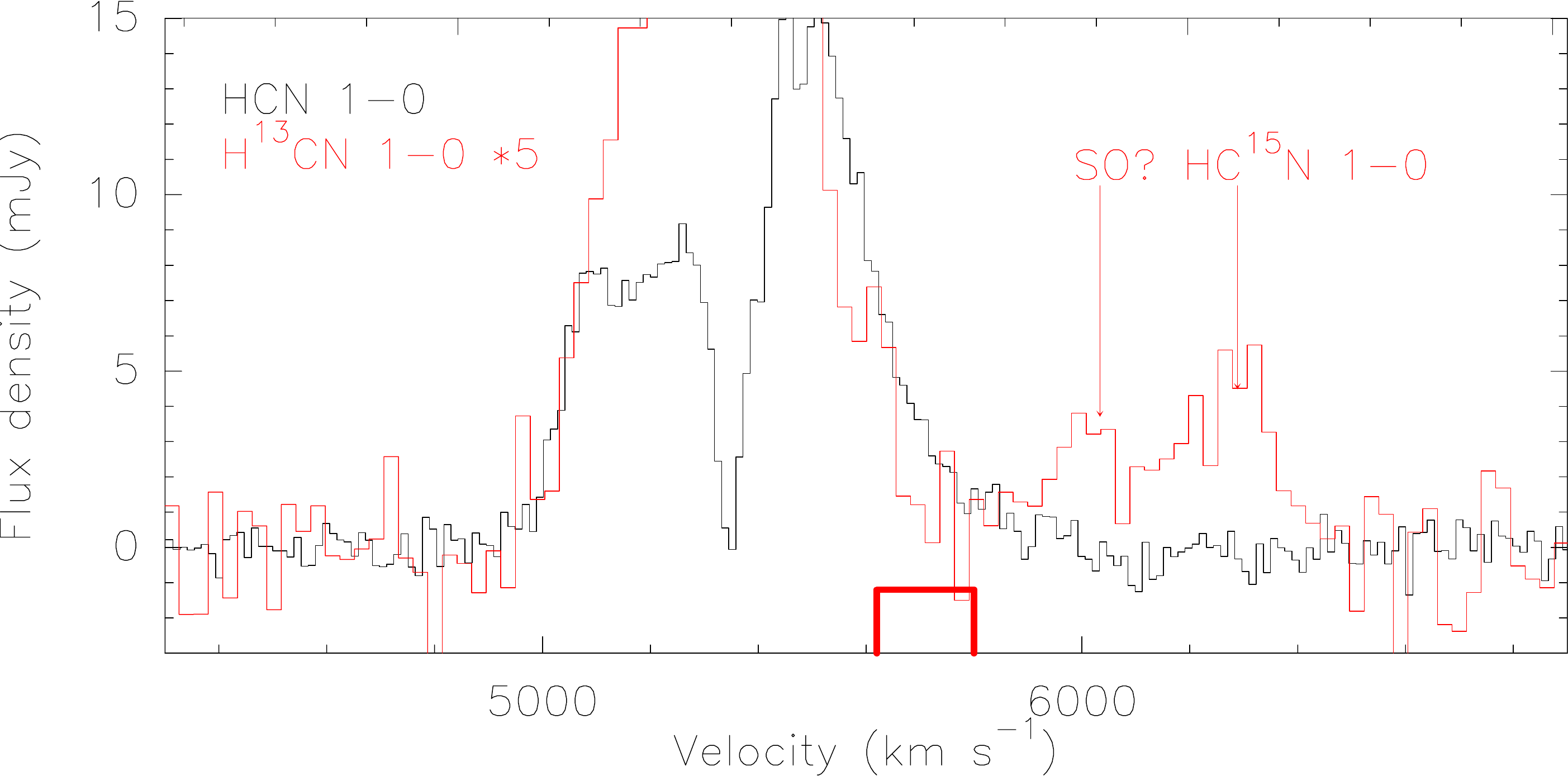}
\includegraphics[angle=0,scale=.25]{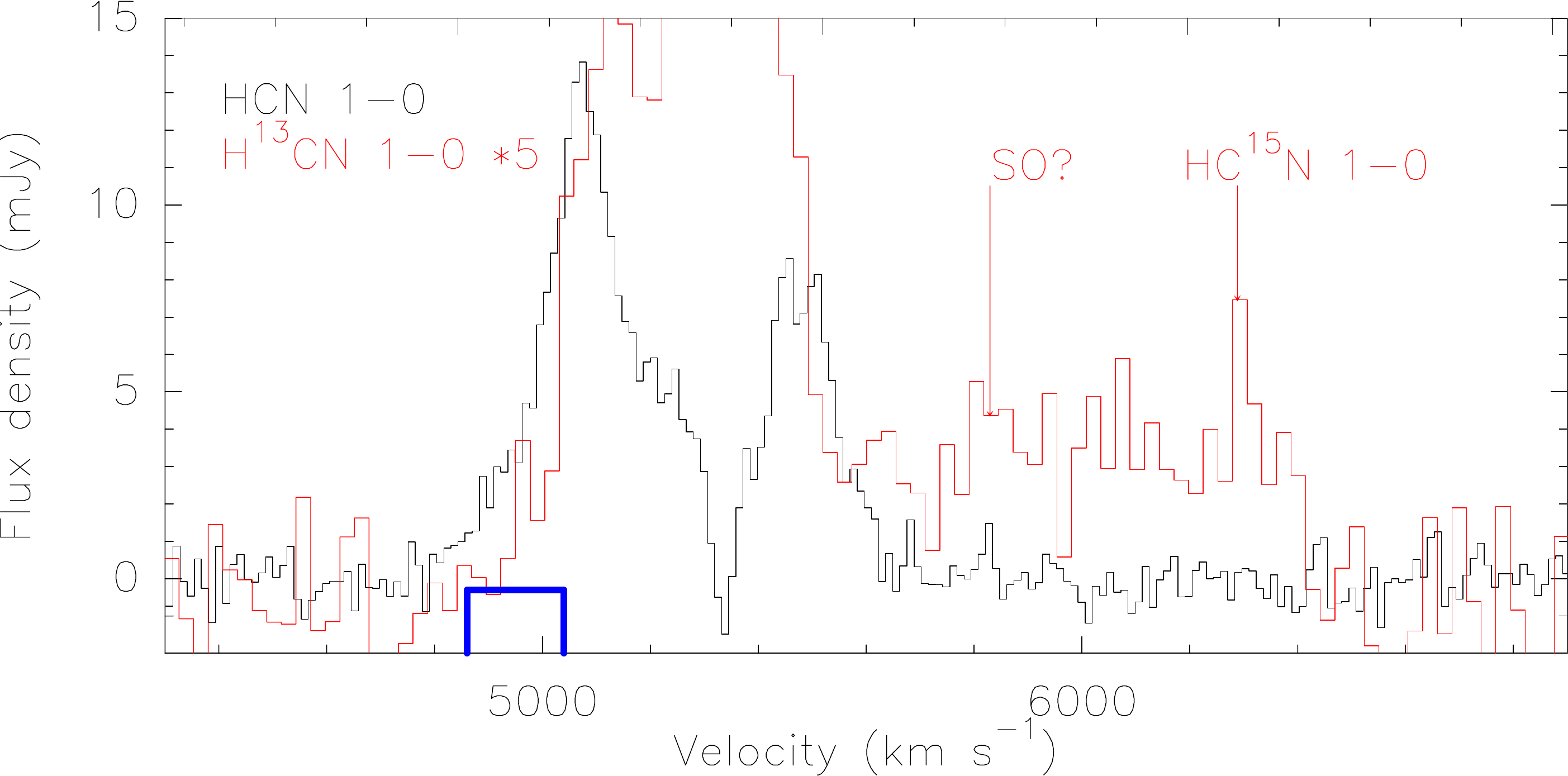}

    \caption{As in Figure 2, but with H$^{13}$CN 1-0  multiplied by 5.   The red window in the top panel is for the velocity range from 5620 to 5800 km s$^{-1}$ which integrates  the red wing of the outflow in Figure 1, while the blue window in the bottom panel is for the blue wing from 4860 to 5040  km s$^{-1}$. }
    \label{fig:figure2}
\end{figure}

\section{Discussion}

\subsection{HCN 1-0  opacity in the outflowing gas}

The average line ratio of HCN/H$^{13}$CN  1-0  in Arp 220 is about five based on single-dish observations \citep{2016MNRAS.455.3986W}, which gives an average opacity of  $\sim$22 for  HCN 1-0, assuming  the  abundance ratio  ($X_{HCN}$/$X_{H^{13}CN}$) is the same as the $^{12}$C/$^{13}$C ratio, namely 100, which was  suggested by \cite{2011A&A...527A..36M}  based on observations of $^{13}$CO/C$^{18}$O 2-1.   The optical depth is obtained with $\frac{\rm{I_{H^{13}CN1-0}}}{\rm{I_{HCN1-0}}}$=$\frac{1-e^{-\tau(\rm{H^{13}CN}1-0)}}{1-e^{-\tau(\rm{HCN}1-0)}}$, where $\frac{\rm{I_{H^{13}CN1-0}}}{\rm{I_{HCN1-0}}}$ is the measured line ratio, while  $\frac{\tau(\rm{H^{13}CN}1-0)}{\tau(\rm{HCN}1-0)}$ is assumed to be the same as the abundance ratio of $^{13}$C/$^{12}$C.
By adopting a HCN/H$^{13}$CN 1-0 ratio of five,  the line profile of H$^{13}$CN is close to that of HCN in the velocity ranges of  5000-5100 km s$^{-1}$ (Figure 3 top)  and 5500-5600 km s$^{-1}$ (Figure 3 bottom). This illustrates that there are no  strong contaminations from HCN 1-0 absorption in the two velocity ranges, and that these ranges are not affected by outflowing gas in the regions with red and blue wings near Arp 220W.
 For the emissions near the central regions of both Arp 220 west and east,  absorption of both HCN 1-0 and H$^{13}$CN  1-0 cannot be neglected, which poses a problem for estimating opacity of HCN 1-0 there.  Absorption of HCN 1-0  towards the continuum emission  or HCN 1-0 emission from inner warm dense molecular gas  can result in overestimation of the   opacity of HCN 1-0 because of the underestimation of HCN 1-0 emission.

The line ratios of HCN/H$^{13}$CN  1-0 in the red and blue wings  are  greater than  9.5  and  9.7, respectively,   giving a 3 $\sigma$   upper limit of $\sim$ 0.1 for the opacity of H$^{13}$CN  1-0 in the outflowing gas.  Assuming an   $X_{HCN}$/$X_{H^{13}CN}$  abundance ratio of 100 as suggested by \cite{2011A&A...527A..36M} with the $^{13}$CO/C$^{18}$O 2-1 line ratio obtained with  the  Submillimeter Array (SMA), the opacity of HCN 1-0 should be less than 10.5.

 The HCN 1-0  flux combined between 5620 and 5800 km s$^{-1}$   in the red wing and  between 4860 and 5040  km s$^{-1}$ in the blue wing   integrated within the dashed cyan  boxes in Figure 1 is 1425.2$\pm$35.7 mJy km s$^{-1}$,  where the flux is from the sum of HCN 1-0 flux from the red wing of 631.1 mJy km s$^{-1}$,  and that from the blue wing of 794.1 mJy km s$^{-1}$, while the noise level is $\sqrt{22.5^2+27.7^2}$ mJy km s$^{-1}$. The corresponding  H$^{13}$CN 1-0 flux is 121.3$\pm$35.7 mJy km s$^{-1}$ derived in the same way, that is, 59.7+61.5 mJy km s$^{-1}$ with noise of $\sqrt{21.1^2+25.9^2}$ mJy km s$^{-1}$.  In other words, H$^{13}$CN 1-0 emission is about 3.4$\sigma$. However, as the low-velocity components  are included, which may be mainly from the disc instead of outflowing gas, the H$^{13}$CN 1-0 emission contribution should be mainly from optically thick non-outflowing gas. Thus, the non-detection of H$^{13}$CN 1-0 emission from the outflowing gas should still be considered and  a 3$\sigma$ upper limit should be used for estimating the line ratio of HCN/H$^{13}$CN  1-0 for the combined regions of outflowing gas, which gives a ratio of greater than 13.3 and an opacity of H$^{13}$CN 1-0 of less than 0.078, or less than 0.11 for both the red and blue wings if calculating them individually. Assuming the  HCN/H$^{13}$CN abundance ratio of 100 as suggested by \cite{2011A&A...527A..36M}, the opacity of  HCN 1-0  should be less than 7.8, which is about one-third of the average value in Arp 220.  Thus, even though we do not have enough information to derive  the opacity of HCN 1-0 in the outflowing gas, the upper limit on HCN 1-0 opacity in the outflowing gas is less than one-third of the average value in Arp 220.

 The $^{12}$C/$^{13}$C ratio in Arp 220 was updated to be 40  with ALMA observations  of $^{12}$CO 3-2 and $^{13}$CO 4-3 \citep{2020ApJ...896...43W}. Therefore, the opacity of HCN 1-0  should be about 40\% of \textcolor[rgb]{0.984314,0.00784314,0.027451}{\textcolor[rgb]{0,0,0}{the opacity estimated using}}  the  $^{12}$C/$^{13}$C ratio of 100  suggested by \cite{2011A&A...527A..36M},  both of which are less than 4.4 as listed in Table 1.  Even with  a 3$\sigma$ upper limit of HCN 1-0 opacity in the outflowing gas  combined for the red and blue components of 3.1, it is still not possible to  draw a firm conclusion as to whether HCN 1-0 is optically thin or thick in the outflowing gas of Arp 220.  The average opacity of   HCN 1-0  in the whole galaxy is down to $\sim$9 using the updated $^{12}$C/$^{13}$C in Arp 220  \citep{2020ApJ...896...43W}.  The optical depth of HCN 1-0 in the outflowing gas is at least three times less than that in the host galaxy of Arp 220 if the same HCN/H$^{13}$CN abundance ratio is used. However,  the possibility of higher HCN/H$^{13}$CN abundance ratio in the outflowing gas than that in the galaxy can also enhance  the  HCN/H$^{13}$CN 1-0 line ratio, even though it is not  likely to lead to significantly different line ratios between the outflowing gas and the gas in the whole galaxy. In summary, even though we can conclude that the optical depth of HCN 1-0 in the outflowing gas is  at least a factor of three lower than the average value for the whole galaxy,  whether HCN 1-0 is optically thick or thin remains uncertain.

\begin{table}
\caption{HCN 1-0 and H$^{13}$CN 1-0  in the  outflows of Arp 220W}             % title of Table
\label{table:1}      % is used to refer this table in the text
\centering                          % used for centering table
\begin{tabular}{c c c c}        % centered columns (4 columns)
\hline\hline                 % inserts double horizontal lines
Line &  Integrated flux  &  Optical depth& \\  
         &  mJy km s$^{-1}$\\  % table heading 
\hline                        % inserts single horizontal line
 HCN 1-0 (blue) &794.1$\pm$27.7&$<$4.4  &  \\      % inserting body of the table
  H$^{13}$CN 1-0 (blue) &$<$83.9 (3$\sigma$) &  $<$0.11   &  \\
  HCN 1-0 (red) &631.1$\pm$22.5& $<$4.4 &  \\      % inserting body of the table
  H$^{13}$CN 1-0 (red) &$<$67.5 (3$\sigma$)  &$<$0.11     &  \\   
\hline                                   %inserts single line
\end{tabular}

Note. The optical depth of  HCN 1-0 is estimated with a $^{12}$C/$^{13}$C ratio of 40 \citep{2020ApJ...896...43W}.
\end{table}

\subsection{Outflow properties and future prospects}

Therefore, even though the accurate opacity of HCN 1-0 of the outflowing gas in Arp 220W  cannot be determined because of the limited sensitivity of the observations,  the non-detection of H$^{13}$CN 1-0  from the outflowing gas provides a good  lower limit  on the HCN/H$^{13}$CN 1-0   line ratio there. Based on this  ratio, an upper limit on the HCN 1-0 opacity of less than one-third of the average value over the whole of Arp 220 can be obtained.  It is necessary to use a smaller conversion factor of  HCN 1-0  luminosity to dense gas mass than that used for normal galaxies, namely of 10   $M_{\odot}$ (K km s$^{-1}$pc$^{2})^{-1}$ \citep{2018ApJ...868...95M}.
 The  conversion factor  under optically thin conditions, namely 0.24 $M_{\odot}$ (K km s$^{-1}$pc$^{2})^{-1}$, was used to  estimate the lower limit of  outflowing mass and mass loss rate in Arp 220W  \citep{2018ApJ...853L..28B}.     The CO 1-0  luminosity of the outflowing gas in Arp 220W is similar to that of HCN 1-0 \citep{2018ApJ...853L..28B}, which means  an even  higher  HCN/CO 1-0 luminosity ratio than that in the massive outflowing gas from the central AGN in Mrk 231 \citep{2012A&A...537A..44A}.  If the same conversion factor of line  luminosity to mass ratio is used in Arp 220W and Mrk 231, the dense gas fraction in the outflowing gas  in Arp 220W is higher than that in Mrk 231, and even the outflowing gas in Mrk 231 is more massive than that in Arp 220W.

However, such  conversion factors  strongly depend on the opacity of CO and HCN lines,  as well as HCN and CO abundances in the outflowing gas.  The abundance issue of HCN when determining dense gas, which can cause uncertainties of estimating the dense gas mass,  can be done with lines of   different  molecules with  high dipole moments, such as HCN, HCO$^+$, HNC, and CS. With observations of different lines of  similar critical density,  the effect of chemical enhancement of special molecules, such as HCN or HCO$^+$,  can be found, as discussed for the outflowing gas  in  Mrk 231 \citep{2016A&A...587A..15L}.  The outflowing gas traced by   HCO$^+$ 1-0  shows similar distribution to that of HCN 1-0 in Arp 220W, with the same dataset \citep{2016PhDT.......253B}, which indicates that astrochemical enhancement of the  abundance of the HCN molecule is not significant in  the outflowing gas.   However, new observations of HNC 1-0 and CS 2-1 are required to further confirm such abundance effects.

The opacity of lines in the outflowing gas can be determined from observations of their optically thin isotopologues, such as H$^{13}$CN 1-0 for HCN 1-0.  The data used in this work were obtained from observations lasting about 1.9 hours on source with ALMA.  If H$^{13}$CN 1-0 data of three times higher sensitivity are required to obtain the  3$\sigma$ upper limit of HCN 1-0 opacity of about  1 with a $^{12}$C/$^{13}$C ratio of approximately 40 measured with ALMA \citep{2020ApJ...896...43W}, the required total telescope time would be about 20 hours; including overheads, this would be prohibitively expensive. Using the RADEX online calculator  \citep{2007A&A...468..627V}, when HCN molecules are close to local thermal equilibrium (LTE),  the  opacity of  HCN 2-1 is about 3.5 times that of HCN 1-0 under  optically thin conditions. Therefore,  it would be  more effective to derive the opacity of  HCN 2-1 with observations of HCN and H$^{13}$CN 2-1 than that of HCN 1-0.    H$^{13}$CN 2-1 in the outflowing gas  in Arp 220W can be expected to be detected within several hours using ALMA, because  HCN 2-1 has moderate opacity of around 0.5 to 1.0, even though  HCN 1-0 is optically thin. Otherwise, an upper limit on HCN 2-1 opacity of approximately 0.5 would also be useful to determine whether or not HCN 1-0 is optically thin.  Stronger HCN 2-1 emission than that of 1-0 was also found in the outflowing gas of Mrk 231 \citep{2016A&A...587A..15L} without significant  line contamination of HCN 2-1.  On the other hand, there is one strong line  within $\pm$700 km s$^{-1}$ range of H$^{13}$CN 2-1, namely HC$_3$N 19-18 at 172.849287GHz at a velocity of $\sim$-300   km s$^{-1}$ relative to H$^{13}$CN 2-1 at  172.67796GHz.  Because of the relatively low abundance ratio of HC$_3$N to HCN and the high density and temperature  requirement for the  excitation of $J$=19,    contamination of HC$_3$N 19-18 at the outflowing region can be neglected.  

Deep $^{13}$CO 2-1 and 1-0 observations may also be useful for determining the opacities  of  CO 1-0 and 2-1 in Arp 220W. However, because the line luminosity of  CO 1-0 is comparable to that of HCN 1-0, there is no great advantage to using  $^{13}$CO lines instead of  H$^{13}$CN.   Combining $^{13}$CO 2-1 and CO 2-1 data with those for HCN 2-1 and H$^{13}$CN 2-1 may be a good choice for determining whether CO and HCN lines in the outflowing gas are optically thick or thin. H$^{13}$CN or $^{13}$CO lines towards local massive AGN molecular outflows or other extreme starbursts with strong outflow,   such as NGC 3256 \citep{2018ApJ...868...95M} or  M82 \citep{2016ApJ...830...72C},  will be helpful for understanding such outflows.

\section{Summary}

 We present an ideal method to derive optical depths of HCN 1-0 in Arp 220 using the HCN/H$^{13}$CN line ratios.  We
apply this method to the
outflowing gas of galaxies   and use it to better determine the conversion factor from
line luminosity to mass. 
With information from spatially resolved  H$^{13}$CN 1-0 and HCN 1-0 observations in Arp 220 with ALMA, we obtain a   3$\sigma$ lower limit on the  HCN/H$^{13}$CN 1-0 line ratio in the outflowing gas of Arp 220W that is at least three times that found for the whole system of Arp 220.  Such a line ratio indicates that the opacity of HCN 1-0 in the outflowing gas of Arp 220W is at least several times lower than that in other regions of Arp 220.  Therefore, a lower  conversion factor  of HCN 1-0 luminosity to dense gas mass should be used  in the outflowing gas  than that used for the whole galaxy of Arp 220. Further sensitive observations  of HCN/H$^{13}$CN 2-1 or 1-0 with ALMA are necessary to constrain the opacity of HCN lines in the outflowing gas in Arp 220W.

%   \begin{enumerate}
%      \item The conditions for the stability of static, radiative
%         layers in gas spheres, as described by Baker's (\citeyear{baker})
%         standard one-zone model, can be expressed as stability
%        equations of state. These stability equations of state depend
%         only on the local thermodynamic state of the layer.
%      \item If the constitutive relations -- equations of state and
%         Rosseland mean opacities -- are specified, the stability
%         equations of state can be evaluated without specifying
%         properties of the layer.
%      \item For solar composition gas the $\kappa$-mechanism is
%         working in the regions of the ice and dust features
%         in the opacities, the $\mathrm{H}_2$ dissociation and the
%         combined H, first He ionization zone, as
%         indicated by vibrational instability. These regions
%         of instability are much larger in extent and degree of
%         instability than the second He ionization zone
%         that drives the Cephe{\"\i}d pulsations.
%   \end{enumerate}

\begin{acknowledgements}
     We thank the anonymous referee for helpful comments to improve the manuscript. This work is supported by  the National Natural Science Foundation of China grant 11988101, 11590783, and U1731237.
This paper makes use of the following ALMA data: ADS$/$JAO.ALMA${\#}$2015.1.00702.S. ALMA is a partnership of ESO (representing its member states), NSF (USA) and NINS (Japan), together with NRC (Canada), MOST and ASIAA (Taiwan), and KASI (Republic of Korea), in cooperation with the Republic of Chile. The Joint ALMA Observatory is operated by ESO, AUI/NRAO and NAOJ.
\end{acknowledgements}

%\begin{Data availability}
%The data used in this paper was from ALMA archive, which can be reached by https://almascience.eso.org/asax/. 
%\end{Data availability}

%\section*{Data availability}

%The data used in this paper was from ALMA archive, which can be reached by https://almascience.eso.org/asax/. 

% WARNING
%-------------------------------------------------------------------
% Please note that we have included the references to the file aa.dem in
% order to compile it, but we ask you to:
%
% - use BibTeX with the regular commands:
%   \bibliographystyle{aa} % style aa.bst
%   \bibliography{Yourfile} % your references Yourfile.bib

\begin{thebibliography}{}

\bibitem[Aalto et al.(2012)]{2012A&A...537A..44A} Aalto, S., Garcia-Burillo, S., Muller, S., et al.\ 2012, \aap, 537, A44. doi:10.1051/0004-6361/201117919
\bibitem[Aalto et al.(2015)]{2015A&A...574A..85A} Aalto, S., Garcia-Burillo, S., Muller, S., et al.\ 2015, \aap, 574, A85. doi:10.1051/0004-6361/201423987
\bibitem[Barcos-Munoz(2016)]{2016PhDT.......253B} Barcos-Munoz, L.\ 2016, Ph.D. Thesis. doi:10.18130/V31S5F
\bibitem[Barcos-Mu{\~n}oz et al.(2018)]{2018ApJ...853L..28B} Barcos-Mu{\~n}oz, L., Aalto, S., Thompson, T.~A., et al.\ 2018, \apjl, 853, L28. doi:10.3847/2041-8213/aaa28d
\bibitem[Chisholm \& Matsushita(2016)]{2016ApJ...830...72C} Chisholm, J. \& Matsushita, S.\ 2016, \apj, 830, 72. doi:10.3847/0004-637X/830/2/72
\bibitem[Cicone et al.(2012)]{2012A&A...543A..99C} Cicone, C., Feruglio, C., Maiolino, R., et al.\ 2012, \aap, 543, A99. doi:10.1051/0004-6361/201218793
\bibitem[Cicone et al.(2014)]{2014A&A...562A..21C} Cicone, C., Maiolino, R., Sturm, E., et al.\ 2014, \aap, 562, A21. doi:10.1051/0004-6361/201322464
\bibitem[Colzi et al.(2018)]{2018A&A...609A.129C} Colzi, L., Fontani, F., Caselli, P., et al.\ 2018, \aap, 609, A129. doi:10.1051/0004-6361/201730576
\bibitem[Feruglio et al.(2010)]{2010A&A...518L.155F} Feruglio, C., Maiolino, R., Piconcelli, E., et al.\ 2010, \aap, 518, L155. doi:10.1051/0004-6361/201015164
\bibitem[Fischer et al.(2010)]{2010A&A...518L..41F} Fischer, J., Sturm, E., Gonz{\'a}lez-Alfonso, E., et al.\ 2010, \aap, 518, L41. doi:10.1051/0004-6361/201014676
\bibitem[Fluetsch et al.(2019)]{2019MNRAS.483.4586F} Fluetsch, A., Maiolino, R., Carniani, S., et al.\ 2019, \mnras, 483, 4586. doi:10.1093/mnras/sty3449
\bibitem[Henkel et al.(2014)]{2014A&A...565A...3H} Henkel, C., Asiri, H., Ao, Y., et al.\ 2014, \aap, 565, A3. doi:10.1051/0004-6361/201322962
\bibitem[Henkel et al.(1998)]{1998A&A...329..443H} Henkel, C., Chin, Y.-N., Mauersberger, R., et al.\ 1998, \aap, 329, 443
\bibitem[Li et al.(2020)]{2020MNRAS.494.1095L} Li, F., Wang, J., Fang, M., et al.\ 2020, \mnras, 494, 1095. doi:10.1093/mnras/staa676
\bibitem[Lindberg et al.(2016)]{2016A&A...587A..15L} Lindberg, J.~E., Aalto, S., Muller, S., et al.\ 2016, \aap, 587, A15. doi:10.1051/0004-6361/201527457
\bibitem[Mart{\'\i}n et al.(2011)]{2011A&A...527A..36M} Mart{\'\i}n, S., Krips, M., Mart{\'\i}n-Pintado, J., et al.\ 2011, \aap, 527, A36. doi:10.1051/0004-6361/201015855
\bibitem[Michiyama et al.(2018)]{2018ApJ...868...95M} Michiyama, T., Iono, D., Sliwa, K., et al.\ 2018, \apj, 868, 95. doi:10.3847/1538-4357/aae82a
\bibitem[Padovani et al.(2011)]{2011A&A...534A..77P} Padovani, M., Walmsley, C.~M., Tafalla, M., et al.\ 2011, \aap, 534, A77. doi:10.1051/0004-6361/201117134
\bibitem[Sturm et al.(2011)]{2011ApJ...733L..16S} Sturm, E., Gonz{\'a}lez-Alfonso, E., Veilleux, S., et al.\ 2011, \apjl, 733, L16. doi:10.1088/2041-8205/733/1/L16
\bibitem[van der Tak et al.(2007)]{2007A&A...468..627V} van der Tak, F.~F.~S., Black, J.~H., Sch{\"o}ier, F.~L., et al.\ 2007, \aap, 468, 627. doi:10.1051/0004-6361:20066820
\bibitem[Wang et al.(2016)]{2016MNRAS.455.3986W} Wang, J., Zhang, Z.-Y., Zhang, J., et al.\ 2016, \mnras, 455, 3986. doi:10.1093/mnras/stv2580
\bibitem[Wheeler et al.(2020)]{2020ApJ...896...43W} Wheeler, J., Glenn, J., Rangwala, N., et al.\ 2020, \apj, 896, 43. doi:10.3847/1538-4357/ab8f32


\end{thebibliography}
%
% - join the .bib files when you upload your source files
%-------------------------------------------------------------------

\end{document}